\documentclass[aps,prl,twocolumn,amssymb,showpacs]{revtex4}
\usepackage{graphicx} 
\newcommand{\de}{\Delta E}
\newcommand{\gi}{G_\infty} 
\newcommand{\ki}{K_\infty} 
\newcommand{\mi}{M_\infty}
\newcommand{\delt}{\Delta E(T)} 
\newcommand{\git}{G_\infty(T)}
 
\newcommand{\mift}{M_\infty(T)}
\newcommand{\ide}{I_{\Delta E}} 
\newcommand{\im}{I_{M_\infty}}
\newcommand{\ig}{I_{G_\infty}} 
\newcommand{\ik}{I_{K_\infty}}
\newcommand{\ms}{\langle x^2\rangle} 
\begin{document} 
\title{Landscape equivalent of the shoving model} 
\author{Jeppe C. Dyre and Niels Boye Olsen}
\affiliation{Department of Mathematics and Physics (IMFUFA), Roskilde
University, POBox 260, DK-4000 Roskilde, Denmark} 
\date{\today}

\begin{abstract}
It is shown that the shoving model expression for the average relaxation time
of viscous liquids follows largely from a classical ``landscape''
estimation of barrier heights from curvature at energy minima. The activation
energy involves both instantaneous bulk and shear moduli, but the bulk modulus
contributes less than 8\% to the temperature dependence of the activation 
energy. This reflects the fact that the physics of the two models are closely 
related.
\end{abstract}
\pacs{64.70.Pf, 62.60.+v, 62.20.Dc}
\maketitle

The physics of highly viscous liquids approaching the calorimetric glass
transition continue to attract attention
\cite{gol64,har76,bra85,ang91,nem95,deb96,dyr98,ang00,don01}. A major
mystery surrounding these liquids is their non-Arrhenius behavior. If
$\tau$ is the average relaxation time and $\tau_0$ the microscopic time (of
order $10^{-13}$s), the 
temperature-dependent activation energy $\delt$ is defined \cite{dyr95,kiv96,note1} by

\begin{equation}\label{1}
\tau(T)\ =\ \tau_0\ \exp\left(\frac{\delt}{k_BT}\right)\,.
\end{equation}
Only few viscous liquids show Arrhenius temperature dependence of the
average relaxation time, i.e., have constant $\delt$. Most liquids have an
activation energy which increases upon cooling.

The ``shoving'' model \cite{dyr96} starts from the standard picture of a viscous
liquid:
At high viscosity almost all molecular motion goes into vibrations around 
potential energy minima. Only rarely do rearrangements take place which move 
molecules from one to another minimum. This view was formulated already by 
Kauzmann in his famous 1948-review \cite{kau48}, and it was the starting point 
of Goldstein's ``potential energy picture'' \cite{gol69} which was recently 
confirmed by computer simulations \cite{hoprefs}. On the short time scale of the 
barrier transition -- expected to last just a few picoseconds -- the surrounding 
liquid behaves as a solid with bulk and shear moduli equal to the instantaneous 
(i.e., high frequency) bulk and shear moduli. Just as in free 
volume theories the shoving model assumes that molecular rearrangements only 
take place when a thermal fluctuation leads to extra space being created 
locally. One may think of the surroundings as being shoved aside, although this 
cause-effect reasoning violates time-reversal symmetry. If there is
spherical symmetry, the surroundings are subject to a pure shear
displacement \cite{dyr98,dyr96,wyl80}. Thus the work done is proportional
to the instantaneous shear modulus $\gi$ and one finds \cite{dyr96} (where $V_c$
is by assumption temperature independent)

\begin{equation}\label{2}
\delt\ =\ \git\ V_c\,.
\end{equation}
This expression \cite{note2} fits data for the non-Arrhenius behavior of 
several molecular liquids \cite{dyr96}; in combination with the
Tool-Narayanaswami formalism the shoving model has been applied also to
structural (i.e., non-linear) relaxations \cite{ols98}. It is
the {\it short-time} shear modulus which appears in Eq. (\ref{2})
because the transition itself is fast (compare, e.g., the analogous
appearance of short-time friction constant in the Grote-Hynes theory 
\cite{gro80}).

The basic assumptions of the shoving model may be summarized as follows
\cite{dyr98}:
\begin{itemize}
\item The main contribution to the activation energy is {\it elastic}
energy.
\item The elastic energy is located in the {\it surroundings} of the
rearranging molecules.
\item The elastic energy is {\it shear} energy.
\end{itemize}
The purpose of this note is to give an alternative justification of Eq.
(\ref{2}) and discuss the interrelation between the two approaches.

First, we review a classical argument estimating the height of the barrier
between two potential energy minima from the curvature around the minima. This
argument is used, e.g., in the Marcus theory for electron transfer reactions
\cite{mar64,cal83}; in 1987 Hall and Wolynes applied this reasoning in their
theory of free energy barriers in glasses \cite{hal87,note3,otherrefs}. Consider 
first a one-dimensional situation with two minima the distance $2a$ apart (Fig. 
1). 
\begin{figure}
\includegraphics[scale=0.45]{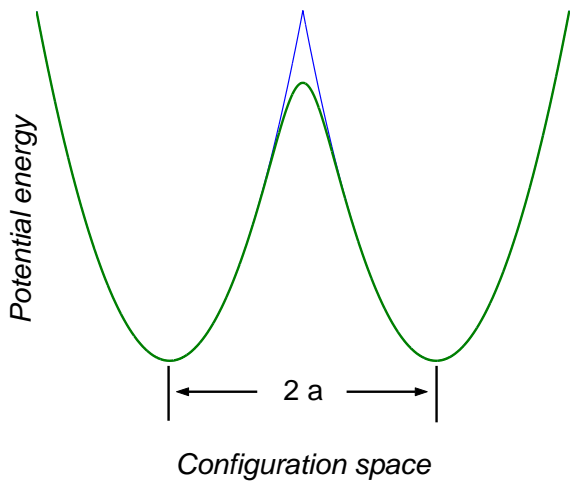}
\caption{Two potential energy minima. The full curve is the potential, the
thin curve is the potential estimated from the curvature at the minima.
According to a classical reasoning used, e.g., in the Marcus theory for
electron transfer reactions, the barrier height may be estimated by the 
parabolas obtained by extrapolating from the minima by second order expansions. 
For simplicity we show the case of identical energy minima; the reasoning 
applies, however, equally for the case of differing minima.
\label{fig1}}
\end{figure}
The thin curve is the potential estimated by second order expansions around
the minima. In this approximation the barrier height is determined
by the curvature of the potential at the minimum: If the potential here is
$U(x)=(\Lambda/2) x^2$, the barrier is given by $\de=(\Lambda/2)a^2$.
Statistical mechanics implies $\ms=k_BT/\Lambda$. Therefore,
comparing different situations with differing potentials {\it but same
distance between minima}, the vibrational mean-square $x$-fluctuation
determines the barrier as follows:

\begin{equation}\label{4}
\delt\ \propto\ \frac{k_BT}{\ms}\,.
\end{equation}
The basic assumptions are that the minima are at fixed distance and that
the parabola approximation gives a good estimate of the barrier
height. Actually, the constant of proportionality in Eq. (\ref{4}) is of no
importance and the true barrier may be consistently smaller than
estimated -- Eq. (\ref{4}) applies as long as the barrier {\it
scales} with that estimated by the parabola approximation.

How does the above reasoning apply to the multidimensional configuration
space where the hills and valleys of the energy landscape live? At low
temperatures a viscous liquid thermally ``populates'' only deep minima in
configuration space. A transition between two deep minima most likely consists 
of a whole sequence of transitions between intermediate shallow minima \cite{bra85}. 
Nevertheless, there is one or more bottlenecks in this sequence. Our 
basic assumption is that at different temperatures the bottleneck transitions 
are the same type of local rearrangements and thus with (virtually) same 
distance between the two minima in configuration space. They occur, however, as 
temperature changes in different surroundings, so the minima involved are 
different and have differing $\ms$. The relevant mean-square $x$-fluctuation is 
to be taken in the direction between the minima. The final assumption needed is 
that this quantity is {\it typical} for the minima, i.e., it is equal to the 
average over all directions. Consequently, the relevant mean-square 
$x$-fluctuation may be evaluated simply as the vibrational mean-square fluctuation 
averaged over all directions. We estimate this quantity by its ensemble 
average over all minima, $\ms$. Note that, since the configuration
space average $\ms$ is equal to the single atom mean square displacement in
three dimensions, this quantity is experimentally accessible in, e.g., neutron 
scattering where it enters into the Debye-Waller factor $\exp(-Q^2\ms)$ (where 
$Q$ is scattering vector) \cite{kit96}. --
Recently, Starr, Sastry, Douglas, and Glotzer arrived at Eq.\ (\ref{4}) from the
free volume perspective \cite{sta02}. Their work includes a direct numerical 
confirmation for a glass-forming polymer melt by calculating the free volume 
$v_f$ and subsequently showing, with $v_f$ as ``mediator,'' that the temperature
dependencies of $\ms$ and $\tau$ are consistent with Eq.\ (\ref{4}).

The next step is to relate $\ms$ to the instantaneous bulk and shear moduli.
On short time scales a  viscous liquid behaves like a solid
\cite{kau48,gol69,hoprefs,dyr99}. In particular, it has well-defined vibrational
eigenstates. We assume that the entire phonon spectrum scales with the long
wavelength limit of the phonon dispersion relation. For a one-dimensional solid
there is only one elastic constant, $C$. The above $\Lambda$ is proportional to
$C$ and consequently $\ms\propto T/C$. In a three dimensional isotropic solid
there are two elastic constants. For each wave vector there are three phonon
degrees of freedom, two transverse and one longitudinal. The relevant transverse
elastic constant is the isothermal ($=$ adiabatic) shear modulus, while the 
relevant longitudinal elastic constant is the isothermal ($\neq$ adiabatic) 
longitudinal modulus $M$ defined \cite{lan70} by $M=K+(4/3)G$
where $K$ and $G$ are bulk and shear moduli.
Averaging over the two types of phonons one finds that the vibrational
mean-square fluctuation is given by $\ms\propto T\left(2/G+1/M\right)$.
This applies for a solid. It applies at short times for a viscous liquid as
well, if $G$ and $M$ are identified with the liquid's instantaneous isothermal
moduli. Thus inserting the expression for $\ms$ into Eq. (\ref{4}) we get

\begin{equation}\label{5}
\frac{1}{\delt}\ \propto\ \frac{2}{\git}\ +\ \frac{1}{\mift}\,.
\end{equation}
Unfortunately, it is not possible to test Eq. (\ref{5}) directly because there 
are no measurements of the instantaneous isothermal bulk modulus.

There are several ways to quantify variations in $\delt$, a liquid's
``fragility''  \cite{ang85}. The standard approach utilizes the
quantity $m$ introduced by Plazek, Ngai, B{\"o}hmer, and Angell
\cite{mdef}:
$m=d\log_{10}(\tau)/d(T_g/T)|_{T=T_g}$ where $T_g$ is the calorimetric
glass transition temperature defined by $\tau(T_g)=10^3$s.
Simple Arrhenius behavior corresponds to $m=16$; most glass-forming liquids have 
fragilities between 50 and 150. As an alternative Tarjus, Kivelson, and 
coworkers proposed to measure the degree of non-Arrhenius behavior at any given 
temperature by the normalized activation
energy: $\delt/\Delta E(T\!\!\rightarrow\!\!\infty)$ \cite{kiv96}. It
is not obvious {\it a priori}, however, that scaling to the high-temperature
limit is physically relevant. Inspired by the Gr{\"u}neisen parameter 
\cite{kit96} and Granato's recent work on interstitialcy relations for the
viscosity \cite{gra02} we suggest that a useful unbiased measure of how much activation energy 
changes with temperature is its logarithmic derivative, $d\ln \delt/d\ln T$. 
Because activation energy increases as $T$ decreases, it is convenient to change 
sign. We thus define the {\it temperature index} $\ide$ of the activation energy 
by

\begin{equation}\label{3}
\ide\ =\ -\ \frac{d\ln\delt}{d\ln T}\,.
\end{equation}
It is straightforward to show that $m=16\big( 1+\ide(T_g)\big)$.

We proceed to express $\ide$ in terms of temperature indices of
instantaneous moduli (same definition). First, note that the 
temperature index $I$ of a sum, $f_1+f_2$, is a 
convex combination of the temperature indices $I_j$ of $f_j$:
$I=\alpha_1I_1+\alpha_2I_2$ where  $\alpha_j=f_j/(f_1+f_2)$. Equation (\ref{5})
thus implies

\begin{equation}\label{6}
\ide\ =\ \frac{2\gi^{-1}}{2\gi^{-1}+\mi^{-1}}\
\ig\ +\  \frac{\mi^{-1}}{2\gi^{-1}+\mi^{-1}}\ \im\,.
\end{equation}
Similarly, $\mi=\ki+(4/3)\gi$ implies

\begin{equation}\label{7}
\im\ =\ \frac{\ki}{\mi}\ \ik\ +\ \frac{4}{3}\frac{\gi}{\mi}\ \ig\,.
\end{equation}
Substituting Eq. (\ref{7}) into Eq. (\ref{6}) leads to

\begin{equation}\label{8}
\ide\ =\ (1-\alpha)\ \ig\ +\ \alpha\ \ik\,,
\end{equation}
where

\begin{equation}\label{9}
\alpha\ =\ \frac{\mi^{-1}}{2\gi^{-1}+\mi^{-1}}\ \frac{\ki}{\mi}\,.
\end{equation}
It is straightforward to show \cite{note4} that one always has

\begin{equation}\label{10}
\alpha\ <\ 0.08\,.
\end{equation}
Thus more than 92\% of the temperature index of the activation energy derives
from the instantaneous shear modulus. The minute influence of the bulk
modulus comes about because of three factors:
\begin{itemize}
\item There are two transverse phonon degrees of freedom, but only one
longitudinal.
\item Longitudinal phonons are associated with a larger elastic
constant than transverse phonons and thus give {\it less} than one
third contribution to the activation energy (Eq.\ (\ref{5})).
\item $\gi$ affects also the longitudinal phonons.
\end{itemize}

We conclude that simplifying \cite{note5}, but not unreasonable,
assumptions in the landscape approach lead to a prediction for the
non-Arrhenius behavior which in practice is going to be hard to distinguish from that of the shoving model. The really interesting question is: What is the
relation between the {\it physics} of the two approaches? At first sight
they seem quite different. There are, however, similarities leading us to
conclude that the physics are actually closely related: The first shoving model
assumption (main contribution to the activation energy is {\it elastic}
energy) is equivalent to the landscape assumption of a parabolic potential
with a curvature which determines the activation energy. The second shoving
model assumption (elastic energy is in the {\it surroundings} of the
reorienting molecules) is consistent with the landscape assumption that the
vibrational mean-square fluctuation is typical, because vibrational
eigenstates in an isotropic solid involve all atoms. And the final shoving
model assumption (elastic energy is mainly {\it shear} energy) is
consistent with Eq.\ (\ref{10}). -- Finally, we would like to remind that
the original shoving model derivation of Eq. (\ref{2}) assumed  spherical
symmetry \cite{dyr96}. In more realistic scenarios there must be some volume
change in the surroundings of the reorienting molecules and thus some
contribution to the activation energy from the instantaneous bulk modulus. While
it is difficult to give absolute bounds on the magnitude of the bulk
contribution, it is noteworthy that Granato -- supported by many others
(see Ref. \cite{gra92} and its references) -- finds that for defect
creation in a crystal the work is overwhelmingly that of a shear
deformation.

\acknowledgments
 This work was supported by the Danish Natural Science Research
Council.


\begin{thebibliography}{99}
\bibitem{note1} We do not distinguish between activation energy and free
energy of activation.
\bibitem{note4} In terms of $k\equiv \ki/\gi$ one has
$\alpha(k)=k/[(k+4/3)(2k+11/3)]$. This function has its maximum at
$k=\sqrt{22}/3$ where $\alpha=0.0794$.
\bibitem{note5} The assumption that the entire phonon spectrum scales with
the macroscopic elastic constants is nontrivial, in particular because of the large number of 
short wavelength phonons. The individual contribution from long wavelength
phonons to $\ms$, however, goes as $k^{-2}$ which compensates for their
small number ($\propto k^2$). Nevertheless, if the short wavelength phonon
contribution to $\ms$ does not scale with the macroscopic elastic
constants, the derivation does not apply. Moreover, in a real viscous
liquid there are localized phonons and our reasoning depends on these
either contributing little to $\ms$ or scaling with the macroscopic elastic
constants.
\end{thebibliography}
\end{document}